\documentclass[aps,prb,twocolumn,superscriptaddress]{revtex4}
\usepackage[dvipdfm]{graphicx}
\usepackage{bm}
\usepackage{amsmath}

\begin{document}


\title{
Scaling behavior of the crossover to short-stack regimes of
Josephson vortex lattices in Bi$_2$Sr$_2$CaCu$_2$O$_{8+\delta}$ stacks}



\author{I. Kakeya}
\email[]{kakeya@kuee.kyoto-u.ac.jp}
\affiliation{Department of Electronic Science and Engineering, Kyoto University,
Nishikyo, Kyoto 615-8510 Japan}
\affiliation{Institute of Materials Science, University of
Tsukuba, Tsukuba, Ibaraki 305-8573 Japan}
\author{Y. Kubo}
\affiliation{Institute of Materials Science, University of
Tsukuba, Tsukuba, Ibaraki 305-8573 Japan}
\author{M. Kohri}
\affiliation{Institute of Materials Science, University of
Tsukuba, Tsukuba, Ibaraki 305-8573 Japan}
\author{M. Iwase}
\affiliation{Institute of Materials Science, University of
Tsukuba, Tsukuba, Ibaraki 305-8573 Japan}
\author{T. Yamamoto}
\affiliation{Institute of Materials Science, University of
Tsukuba, Tsukuba, Ibaraki 305-8573 Japan}
\author{K. Kadowaki}
\affiliation{Institute of Materials Science, University of
Tsukuba, Tsukuba, Ibaraki 305-8573 Japan}


\date{\today}
\begin{abstract}
We report the systematic investigations of the oscillation of the Josephson vortex (JV) flow resistance
in Bi$_2$Sr$_2$CaCu$_2$O$_{8+\delta}$ micro-fabricated junctions with various geometries
and superconducting anisotropy parameters.
As the applied magnetic field parallel to the $ab$-plane is increased,
oscillation with a period corresponding to a $\phi_0/2$ par atomic Josephson junction
changes to oscillation with a doubled period.
This crossover is scaled by both the junction length and the anisotropy parameter,
indicating that the bulk inductive coupling that favors the triangular JV lattice is replaced
with the surface deformation energy as the dominant interaction for a JV lattice.
These results suggest that the in-phase square JV lattice is pronounced at a higher magnetic field
in a smaller and more anisotropic sample.
\end{abstract}


\maketitle


Highly anisotropic high-$T_c$ superconductors (HTSC) such as Bi$_2$Sr$_2$CaCu$_2$O$_{8+\delta}$ (Bi2212)
are well described as stacks of weakly coupled Josephson junctions (JJ's) that consist of CuO$_2$
superconducting and charge reservoir block layers,\cite{Kleiner:1992,Oya:1992} referred to as
intrinsic Josephson junctions (IJJ's). Not only the inductive coupling~\cite{Sakai:1993} but also
the capacitive coupling~\cite{Koyama:1996} between the stacked Josephson junctions are responsible
for such peculiar properties of IJJ's as multiple branches in the current-voltage ($IV$) characteristics
and the longitudinal Josephson plasma mode.\cite{PRB98} Rich collective excitation phenomena
attributed to these couplings among identical stacked junctions\cite{Yurgens:2000}
can be adopted for investigating physics of HTSC and
also promise us potentials for future device applications.

In the magnetic field parallel to the $ab$ plane,
where Josephson vortices (JV's) quantized as a unit of $\phi_0=h/2e$ are induced,
various static and dynamic arrangements of JV's are expected.~\cite{Bul92a,Hu_:2000}
The variety of phenomena is given by the couplings between the adjacent JJ's,
which make IJJ's a sharp contrast with artificially stacked JJ's.
The other characteristic features of the JV system in parallel
magnetic fields have been found in the $IV$
characteristics,~\cite{Hec:1997,Krasnov:1999} the Josephson plasma
resonance,~\cite{PRB05} and etc.

Recently, Ooi {\it et al.}~\cite{Ooi:2002} found a periodic oscillation of the JV flow resistance
with a period of $\phi_0/2$ in Bi2212 IJJs with widths of 20-50 $\mu$m as a
function of the magnetic field. They interpreted this is due to the coherent flow of the triangular
JV lattice. Subsequently, Koshelev~\cite{Koshelev:2002} and Machida~\cite{Mac:2003}
explained the oscillation by calculating the JV flow resistance by taking the surface
barrier effect into consideration. In this model, since the surface barrier introduced by the boundary condition at
the junction edges only influences the positions of JV's close to the surfaces,
a bulk property (triangular JV lattice) is probed by the surface current.
Reducing the junction width to $\sim \mu m$ is expected to
make the surface barrier effect compete with the bulk inductive coupling between the JV
arrays in neighboring layers and result in drastically different phenomena.
In smaller junctions $< 10 \mu m$, the authors found oscillations with the doubled period $\phi_0$ at high magnetic fields,~\cite{LI-1}
and Hatano et al., reported the $\phi_0$ oscillation in a sub-micron junction.\cite{Urayama:2006}
Motivated by our works, Machida\cite{Machida:2006} and Koshelev\cite{Koshelev:2007a} proposed advanced calculations that the triangular and the
square JV lattices appear alternatively as a function of magnetic field and that the square
lattice is more favorable for higher magnetic field and in samples with smaller widths.
Since the square JV lattice is a result of in-phase locked oscillation of the Josephson currents of the stacked IJJ's,
realization of a square JV structure has been considered as a central issue for an electromagnetic radiation from IJJ.
Although recently reported strong terahertz emission from IJJ's in the absence of applied magnetic field~\cite{Ozyuzer:2007,Rhodes_kadowaki}
is attributed to the size-restricted cavity resonance of the rather large mesas ($\sim 100 \mu$m),
tuning excited wavelength by JV lattice like a field tunable photonic crystal \cite{PRB05,Savelev:2006} is a next goal of the IJJ emitter.

In this paper, we present the experimental results of the JV flow resistance and $IV$ characteristics in
Bi2212 micro-structured single crystals with dimensions of 1$-$6 $\mu$m more clearly and systematically
than in the previous work.\cite{LI-1}
In the smaller samples and
at higher fields, we found periodic oscillation of the JV flow resistance with the period of the
$\phi_0$ instead of $\phi_0/2$ as argued by the theories.
The results clearly indicate that the crossover of the two oscillations is well tuned by
both the superconducting anisotropy parameter and the junction size,
which may prove formation of the square JV lattice in the static and slowly moving dynamic states.

Bi2212 single crystals grown with the traveling floating method were fabricated into bridge-shaped mesoscopic junctions
with a focused ion beam (FIB) machine SMI2050MS (SII NanoTechnology Inc.).~\cite{Kim:1999} The
fabrication details are described in Ref. \cite{Rhodes_kakeya}.
Here, we report the results of seven junctions obtained from different slightly over-doped crystals
as listed in Table \ref{tab1}. $T_c$'s are in the range of 84 -- 90 K.
External magnetic fields were applied by split pair superconducting magnets that generate a
horizontal magnetic field up to 80 kOe. Rotating the magnetic field around the $ab$-plane with a high
precision rotator, $c$-axis resistance suddenly increases in the vicinity of the $ab$-plane.\cite{LI-1} This
resistance is attributed to the Josephson vortex flow resistance, and the lock-in state without
pancake vortices is realized in this angle range. We set the field alignment parallel to the
$ab$-plane with an accuracy of 0.01 degree.

\begin{table}[bt]
\caption{\label{tab1}List of samples.
$L$, $W$, and $t$ denote length ($\perp \bm{H}$), width ($\parallel \bm{H}$), and thickness
($\parallel c$) of junctions, respectively. Oscillation starting field $H_s$ and crossover field
$H_{1/2}$ were directly given by measurements. $\gamma$ was derived from $H_{1/2}$. Units for
dimensions and magnetic fields are $\mu$m and kOe, respectively.}
\begin{ruledtabular}
\begin{tabular}{lllllllll}
 & $L$ & $W$ & $t$ &  $H_p$ & $H_s$ & $H_{1/2}$ & $\gamma$  \\
\hline
B19 & 1.9 & 11.7 & 0.5  & 6.7 & 15 & 16 & 119 \\
E44 & 4.4 &  9.5 & 0.15 & 3.04 & 6.0 & 34 & 130\\
H41 & 4.1 &  5.5 & 2.0 & 3.24 & 7.0 & 42 & 114 \\
H55 & 5.5 &  4.1 & 2.0 & 2.49 & 5.3 & 35 & 142 \\
P52 & 5.2 & 3.5 & 0.36 & 2.43 & 5.0 & 30 & 155  \\
Q50 & 5.0 & 4.9 & 0.16 & 2.60 & 5.3 & 46 & 121  \\
V31 & 3.1 & 5.1 & 1.0 & 4.11 & 7.6 & 33 & 173  \\
\end{tabular}
\end{ruledtabular}
\end{table}


\begin{figure}
\begin{center}
\includegraphics[width=\linewidth]{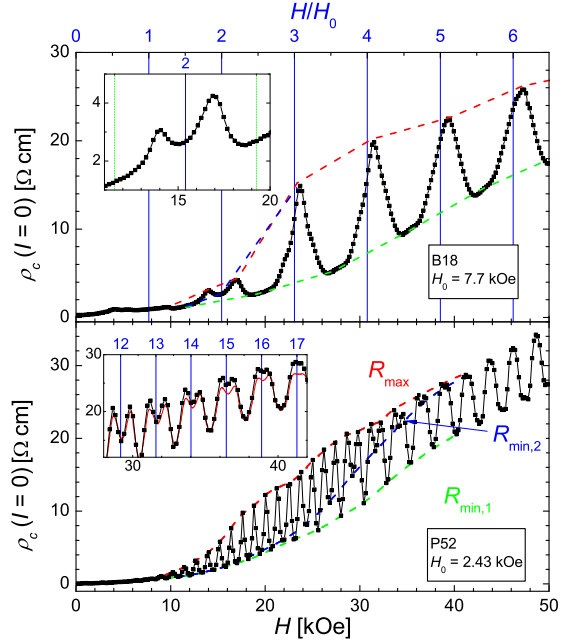}
\end{center}
\caption{(Color online)Field dependence of differential $c$-axis resistivity in B19 (a)
and P52 (b) at 60 K.
Solid symbols represent differential resistance $dV/dI$ obtained from $IV$ characteristics.
Broken curves connect maxima and minima points in $\rho(H)$.
In the inset of (b), solid curve indicates ohmic resistances $V/I$ obtained by sweeping magnetic field at $I=1.2\mu$A.
Blue vertical grids in (a) and inset of (b) indicate that the fields of $H/H_0$ are integers.
} \label{R-H}
\end{figure}

Figures \ref{R-H}(a) and (b) represent the magnetic field dependencies of zero-current differential
$c$-axis resistivity $\rho_c=LW/t \cdot dV/dI|_{I=0}$ at 60 K in sample B19 and P52, respectively.
The data were derived from $IV$ characteristics measured under external
magnetic fields parallel to the $ab$-plane changed step by step.
This method for evaluation of the JV feature is less ambiguous than the method with the ohmic resistance $V/I$.
The $\rho_c$ in P52 smoothly
increases below 7 kOe with increasing field from zero and then begins to oscillate with an oscillation
center monotonically increasing, as shown in Fig. \ref{R-H}(b). Oscillation period $H_p$ is 1.2
kOe, which approximately coincides with adding a {\it half} flux quantum $\phi_0/2$ to a JV array
in the block layer as $H_0/2=\phi_0/2Ls=$ 1.3 kOe for P52, where $s=15$ \AA  is the periodicity of
the CuO$_2$ double layers of Bi2212. This oscillation is quite similar to the previous results.~\cite{Ooi:2002}
The boundary effect for the triangular JV lattice is considered to be enhanced when the number of JV's in
a layer corresponds to either integer ($H/H_0=k$; full-integer matching) or half odd integer($H/H_0=k+1/2$; half-integer matching)
due to matching of the periodicity of the JV lattice with the junction width for both two cases. These two matchings
yield the oscillation with period of $H_0/2$ and the bottom at $H/H_0=k$ and $k+1/2$.

With increasing magnetic field, the full-integer matching was diminished above 25 kOe and finally
disappears above 40 kOe, while the half-integer matching does not change to 50 kOe. As a
result, $H_p$ becomes 2.43 kOe, which is twice the value of the $H_p$ below 35 kOe and close to the $H_0$ value of
2.50 kOe. This crossover from $\phi_0/2$ to $\phi_0$ oscillation was observed in a lower field
region in a sample with smaller $L$. Figure \ref{R-H}(a) shows a $\rho_c-H$ curve in B19 with
$L=1.9$ $\mu$m. $\phi_0$ oscillation was certainly observed above 20 kOe, and only a small dip at
$H/H_0=2$ was found as a remnant of the $\phi_0/2$ oscillation, as shown in the inset of Fig.
\ref{R-H}(a).

\begin{figure}
\begin{center}
\includegraphics[width=0.8\linewidth]{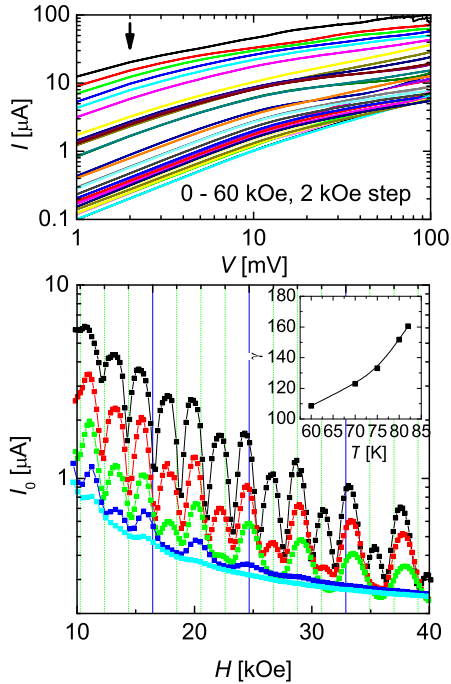}
\end{center}
\caption{(Color online)(a) $IV$ characteristics at 60 K in V31 ($T_c=$ 84 K) at various magnetic fields from 0 (top) to 60 kOe (bottom).
(b) Field dependence of $I_1$, current at $V=2$mV (the arrow in (a)), at 60, 75, 80, and 82 K from top to bottom.
Inset shows temperature dependence of anisotropy parameter $\gamma$ obtained by Eq. (\ref{B_1/2}).
\label{Jc-H@V31}}
\end{figure}

Since linear $IV$ characteristics were observed even in the low current limit
as displayed in Fig. \ref{Jc-H@V31}(a), it was implied that there is no true critical current in IJJ with JV lattices.
This makes a remarkable contrast with an ordinary vortex system under the $c$-axis magnetic field for HTSC,
where a super-linear increase in voltage in $IV$ characteristics is regarded as the critical current.
In order to compare with theoretically-derived field dependencies of critical current density $J_c(H)$,
the current $I_0(H)$ at which the induced voltages along the junction are 2 mV is plotted as a function of magnetic field at various temperatures
in Fig. \ref{Jc-H@V31}(b).
At 75 K below 15 kOe, $I_0(H)$ oscillates with a period of $H_0/2$ as proposed in Ref. \cite{Koshelev:2002} for $J_c(H)$,
while $I_0(H)$ above 35 kOe shows oscillation with the period being $H_0$ as well as $\rho_c(H)$.
The behavior of $I_0(H)$ in the high field region turns to be similar to
$J_c(H)$ in a single Josephson junction (Fraunhofer pattern) that has local maxima at fields
corresponding to $(k+1/2)\phi_0$. Such a Fraunhofer oscillation in $J_c(H)$ of multilayered system
like Bi2212 is expected when condition $L \ll \lambda_J \equiv \gamma s$ is
satisfied.~\cite{Bul92a}
However, $\lambda_J$ is estimated to be approximately 0.2 $\mu$m for
$\gamma = 100-200$, which is much smaller than the $L$ of these samples.

The experimental results demonstrate that the $\phi_0$ oscillation is a phenomenon induced by the external magnetic fields,
thus it is important to consider competition between the bulk and the boundary effects for JV's.
At relatively low fields where JV's start to form dense triangular
lattices, the JV positions are mostly determined by the sharing interaction between vortices in
adjacent layers due to the inductive coupling between layers. With increasing magnetic field,
vortices inside the layer become closer, and interaction between intra-layer JV's is pronounced. The
JV system is finally dominated by the surface deformation energy that lets JV's align in phase and
turns toward the square lattice in the high field region at the cost of sharing energy between JV
arrays. Since the sharing energy of JV's in a smaller junction with less JV's is smaller while the surface deformation does not depend on $L$,
the $\phi_0$ oscillation is observed at a lower field.
The single-junction-like behaviors are more pronounced in samples with shorter $L$ and at higher $H$.~\cite{Koshelev:2002}

The crossover between the triangular and the square JV lattices is quantitatively treated by Koshelev.~\cite{Koshelev:2007a}
He derived the amplitudes of the resistance oscillation due to the half-integer matching and
 the full-integer matching, $\delta R_1$ and $\delta R_2$, as functions of magnetic field.
The magnetic field at $\delta R_2/\delta R_1=1/2$ is given by the junction length $L$ and the anisotropy parameter $\gamma$ as:
\begin{equation}
H_{1/2}=1.302\frac{\phi_0 L}{2 \pi \gamma^2 s^3}.
\label{B_1/2}
\end{equation}
In order to extract generic boundary effect from our results in many samples with various oxygen concentrations,
the sample specific parameter $\gamma$ should be taken from measurable parameters.
The comparison between the $ab$-plane
resistance before FIB fabrication the $c$-axis resistance~\cite{Yu:2007}
does not work in this case because oxygen would be reduced during FIB fabrication in the vacuum and
implanted gallium ions may damage the junction. These influences are more serious in junctions with
widths less than 5 microns. To obtain the anisotropy parameter in the exactly same situation as the
sample, we use the formation of the dense triangular JV lattice,
which is considered necessary for $\phi_0/2$ oscillation. Using a minimum field for
triangular dense JV lattice $H_{tr}=1.33 \phi_0/2 \pi \gamma s^2$,\cite{Nonomura:2006}
we obtain a simple relation:
\begin{equation}
H_{1/2}/H_{tr}^2=0.736 \frac{2\pi s L}{\phi_0}  = 3.35 [1/\mu m{\rm T}] L [\mu m],
\label{reduced}
\end{equation}
where the magnetic fields are given in the unit of Tesla.
Assuming that $H_{tr}$ corresponds to $H_s$, the bare boundary effect for JVs may be extracted from the results.

\begin{figure}[bt]
\begin{center}
\includegraphics[width=1.0\linewidth]{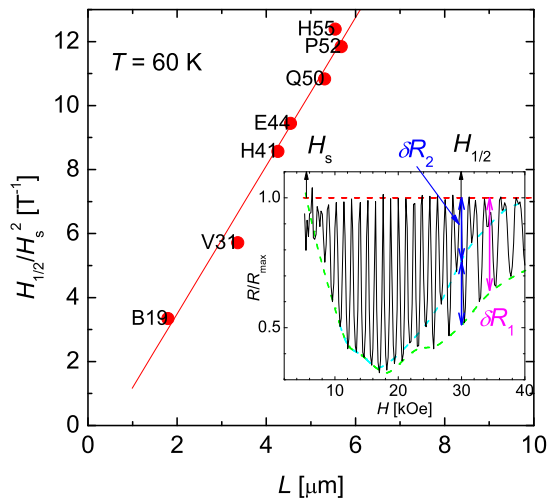}
\end{center}
\caption{(Color online)$L$ dependence of crossover field $H_{1/2}$ in which $\gamma$ is renormalized.
Inset: Method to extract $H_s$ and $H_{1/2}$. Solid and broken curves are those in Fig. \ref{R-H}(b)
normalized by $R_{max}$ (red (dark gray) broken curve).
\label{norm_Hth}
}
\end{figure}

Figure \ref{norm_Hth} shows the values of $H_{1/2}/H_s^2$ as a function of junction length $L$.
$H_{1/2}/H_s^2$ is roughly proportional to $L$. $H_{1/2}$ and $H_s$ were obtained by extracting
$\delta R_1$ and $\delta R_2$ from the resistance minima and maxima, as shown in the inset of Fig.
\ref{norm_Hth}. $\delta R_1$ and $\delta R_2$ are given by subtracting $R_{min,1}$, the smooth curves connecting
resistance minima at $H/H_0=k+1/2$, and $R_{min,2}$ from $R_{max}$ in $\rho_c(H)$, respectively.
Since it is difficult to decide the onset of
the oscillation, $H_s$ was defined as $\delta R_1=0$ by extrapolating
$R_{min,1}/R_{max}$ to 1, so that $H_s$ may include more ambiguity than $H_{1/2}$. The slope of the
fitted line is slightly smaller than Eq. (\ref{reduced}) with a factor of 0.68, meaning $H_s=1.2H_{tr}$.
Below $H_{tr}$ JV lattices incommensurate along the $c$ axis are expected in the bulk
as results of the sharing transitions from the triangular JV lattice.~\cite{Nonomura:2006}
Considering the boundary effect which may suppress the sharing transition, the deviation could be reasonable.
The result that $H_{1/2}$ can be scaled by
$\gamma$ suggests that the crossover between $\phi_0/2$ and $\phi_0$ oscillations is a result of
competition between bulk sharing energy and the edge boundary effect for JVs. The argument that the
$\phi_0$ oscillation is attributed to the Fiske step at higher bias current~\cite{Ustinov:2005} is not
supported because $dV/dI$ at $I=0$ shows almost same behavior as ohmic resistance $V/I$, as
shown in the inset of Fig. \ref{R-H}(b).
Therefore such a non-linear effect in voltage should not be considered in this current region.

The temperature dependence of the crossover field is also consistent with the preceding discussions.
As temperature is increased, $H_{1/2}$ considerably shifts to a lower field from 60 to 82 K
as shown in Fig. \ref{Jc-H@V31}. We derived $\gamma$ with Eq. (\ref{B_1/2}) and
plotted it in the inset of Fig. \ref{Jc-H@V31}(b) as a function of temperature. It has been reported
that $\gamma$ drastically increases in the vicinity of $T_c$.~\cite{Mirkovic:2002,Kon:2006}
It was also found that the amplitude of the oscillation of both $\rho_c(H)$ and $I_0(H)$
is maximized at 60-70~K. This can be understood that the synchronization of stacked junctions is
optimized at the temperatures: thermal fluctuation below 60 K smears the inhomogeneities contained
in the crystal and helps form regular JV lattice, but thermal fluctuation above 70 K smears the
matching conditions.



In summary, we investigated the Josephson vortex lattice structure in mesoscopic ($\sim\mu$m)
IJJ's with a probe of the JV flow resistance along the $c$ axis. The $c$-axis resistivity
$\rho_c$ oscillates as a function of applied magnetic field with a oscillation period changing from
$\phi_0/2$ to $\phi_0$, which indicates triangular JV lattice and a possible square JV lattice,
respectively. Crossover field $H_{1/2}$ linearly decreases with junction length $L$. This is
attributed to the interplay between the boundary effect and the inductive coupling between
JV arrays in IJJ's.

\begin{acknowledgments}
We thank M. Machida and A. Koshelev for their fruitful discussions and critical comments. This work
was partially supported by the Grant-in-Aids for Young
Scientists (B) and the Core-to-Core Program ``Nanoscience and
Engineering in Superconductivity'' by Japan Society for the Promotion of Science (JSPS).
\end{acknowledgments}

\bibliography{MyNewBibs,MyNewPubs}

\end{document}